\documentclass[sigconf,screen]{acmart}

\AtBeginDocument{%
  }

\copyrightyear{2026}
\acmYear{2026}
\setcopyright{cc}
\setcctype{by}
\acmConference[CHI EA '26]{Extended Abstracts of the 2026 CHI Conference on Human Factors in Computing Systems}{April 13--17, 2026}{Barcelona, Spain}
\acmBooktitle{Extended Abstracts of the 2026 CHI Conference on Human Factors in Computing Systems (CHI EA '26), April 13--17, 2026, Barcelona, Spain}
\acmDOI{10.1145/3772363.3799270}
\acmISBN{979-8-4007-2281-3/2026/04}

\begin{document}

\title[PlantWhisperer: Designing Conversational AI to Support Plant Care]{PlantWhisperer:\\Designing Conversational AI to Support Plant Care}

\author{Daniel Mejer Christensen}
  \authornote{Authors contributed equally to this research.}
\orcid{0009-0003-0467-8556}
\email{dmch23@student.aau.dk}
\affiliation{%
  \institution{Aalborg University}
  \city{Aalborg}
  \country{Denmark}
}

\author{Katja Stougård Jørgensen}
\authornotemark[1]
\orcid{0009-0002-7564-9317}
\email{kjorge22@student.aau.dk}
\affiliation{%
 \institution{Aalborg University}
 \city{Aalborg}
 \country{Denmark}
}

\author{Josefine Palsgaard Wyrtz}
\authornotemark[1]
\orcid{0009-0007-7874-835X}
\email{jwyrtz23@student.aau.dk}
\affiliation{%
  \institution{Aalborg University}
  \city{Aalborg}
  \country{Denmark}
}

\author{Jennie Torp Overgaard}
\authornotemark[1]
\orcid{0009-0009-2912-7658}
\email{joverg23@student.aau.dk}
\affiliation{%
  \institution{Aalborg University}
  \city{Aalborg}
  \country{Denmark}
}

\author{Niels van Berkel}
\email{nielsvanberkel@cs.aau.dk}
\orcid{0000-0001-5106-7692}
\affiliation{%
  \institution{Aalborg University}
  \city{Aalborg}
  \country{Denmark}
}

\author{Joel Wester}
  \authornote{Corresponding author.}
\email{joel.wester@di.ku.dk}
\orcid{0000-0001-6332-9493}
\affiliation{%
  \institution{University of Copenhagen}
  \city{Copenhagen}
  \country{Denmark}
}
\affiliation{%
  \institution{Aalborg University}
  \city{Aalborg}
  \country{Denmark}
}

\renewcommand{\shortauthors}{Mejer Christensen et al.}

\begin{abstract}
Research in Human-Computer Interaction (HCI) has shown that caring for others, including both humans (e.g., close friends) and computers (e.g., Tamagotchi), can have a positive effect on people’s wellbeing.
However, we know less about the potential role of conversational AI in such settings.
In this work, we explore how AI chatbots can support plant care and, in turn, positively influence people's well-being. 
We developed a mobile application that allows users to `talk' to their plants via chatbots. 
We evaluated the application with ten participants and conducted semi-structured interviews based on Seligman’s PERMA model, which identifies pillars of psychological well-being.
Our findings suggest positive effects, with participants reflecting on a sense of connection to their plants and corresponding feelings of accomplishment. 
While our findings suggest that participants were generally positive about the app, they also raised concerns about the diverse preferences and expectations of users regarding interactions with chatbots representing plants. 
\end{abstract}

    \begin{teaserfigure}
 \centering
\includegraphics[width=0.88\textwidth]{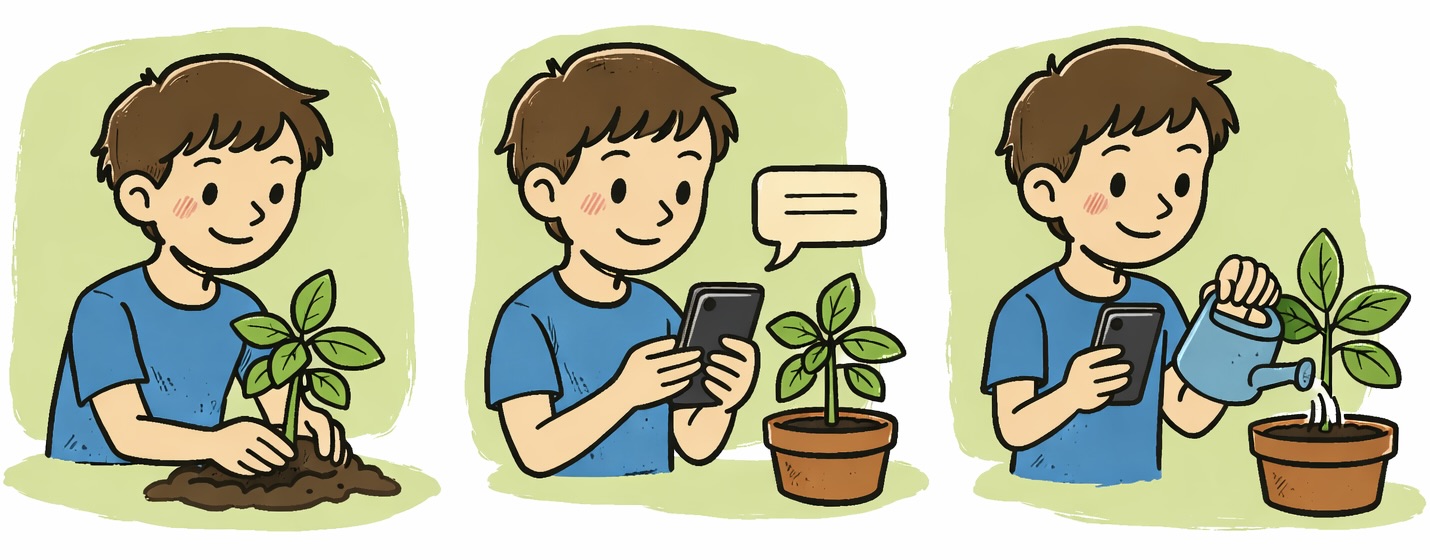}
\caption{PlantWhisperer is an AI-powered mobile application designed to help users care for their plants. 
Users can manually add their plants and receive personalised feedback about each one.
Specifically, they chat with plant characters represented by LLM-driven chatbots that communicate their distinct characteristics.}
\Description{Figure illustrates an individual caring for the plant; then chatting with the plant for plant care support; and then watering the plant based on the guidance provided.}
\label{fig:illustration}
\end{teaserfigure}


\maketitle

\section{Introduction} 

Previous studies have emphasised the positive impact of horticultural activities on well-being~\cite{10.1145/3715336.3735689, 4132d5d497394a42ac5b0af99df406c3, 10.1145/3689050.3705996}.
A recent review of 40 studies also suggests that gardening has a beneficial effect on mental health and quality of life~\cite{Pantiru2024}. 
In the medical field, horticulture therapy (HT) uses plant-based activities to promote physical and mental well-being, and gardening can enhance self-esteem and self-efficacy~\cite{gardeningSelfEsteem, horticultureTherapy}. 
Horticultural practices in households can take the form of kitchen gardens, which are suggested as a sustainable way to support psychological and physical health~\cite{Galhena2013, CHALMINPUI2021103118}.
However, despite the positive effects of horticultural activities on well-being, individuals may still find it challenging to initiate or sustain them.

People also feel better when they care for others, including through chatbots~\cite{10.1145/3290605.3300932}. 
Conversational AI is receiving increased attention for its ability to provide personalised chatting experiences, form emotional bonds with users~\cite{ha2024clochatunderstandingpeoplecustomize}, or influence their well-being in other ways.  
For example, a recent study found that interacting with AI representing animals positively influenced participants' sustainable intentions, reflections, and choices about plastic pollution by using character narratives to boost emotional engagement, perceived empathy, and anthropomorphism~\cite{pataranutaporn2025oceanchateffectvirtualconversational}. 
Another example is CoRoot, a collaborative planting system designed to help family members build closer intergenerational relationships and share gardening knowledge and skills by growing plants together~\cite{xu2024coroot}.
Such emerging interactions provide an interesting research direction, in which interactions with AI can promote emotional engagement and empathy~\cite{pataranutaporn2025oceanchateffectvirtualconversational}. 
Nevertheless, little is known about how horticultural activities can be supported and mediated through AI chatbots, or what potential benefits such an application might offer.

Therefore, we explored how people perceive an LLM-powered chatbot `representing' plants in the context of kitchen gardening.
Specifically, we implemented and deployed a mobile application: \textit{PlantWhisperer}. 
The primary function of the chatbot is to represent kitchen plants through natural language and distinct characteristics, enabling users to interact with them in an engaging and personalised manner.
We conducted a pilot study involving four participants to inform and iterate on the design of \textit{PlantWhisperer}.
The pilot study revealed that chat messages should be concise and that both context and distinct plant characteristics were important to users' overall experience.
Our main evaluation revealed key insights regarding \textit{PlantWhisperer's} potential to support well-being.
Participants reported appreciating the chatbots' emotionally expressive interactions, with several noting that the tone contributed to a stronger sense of connection with the plant. 
We also noted that some users preferred a more direct interaction style, highlighting the importance of customising tone, length, and interaction style to suit individual needs. 

One takeaway we emphasise is the emotional expressiveness of plant chatbots that can foster a connection with plants through engagement.
Our work highlights how using AI chatbots to mediate interactions with real-world plants can positively benefit people's well-being and opens up interesting venues for future work.

\section{Plant Care, Well-Being, and Conversational Technologies}
Mental health research has suggested that gardening is increasingly being seen as a helpful tool for supporting mental health~\cite{ClatworthyJane2013Gaam}. 
In the UK, the number of gardening projects for people with mental health issues has grown from 45 in the 1980's to over 900 by 2005. 
This aligns with the wider narrative that nature can boost well-being, with many claims that time spent in natural settings can have psychological benefits. 
This is supported by research that reviewed gardening projects, which found that gardening can lead to fewer symptoms, better social connections, and new skills for people dealing with mental health challenges \cite{ClatworthyJane2013Gaam}. 
Related, HCI work on chatbots for self-compassion 
explored two versions of a chatbot, Vincent, with one as care-giving and the other as care-receiving \cite{10.1145/3290605.3300932}.
The results of testing Vincent indicate that the care-receiving chatbot had a greater impact on participants than the care-giving chatbot. The article emphasises that practising compassion not only helps offer support to others but also enhances the ability to apply it to oneself \cite{10.1145/3290605.3300932}. 

Interaction between people and plants can be designed in a personal and engaging way through conversation. Conversational interfaces have been shown to support meaningful interactions by allowing users to communicate through familiar, text-based communication ~\cite{klopfenstein2017theriseofbots}.
The recent development and accessibility of LLMs, such as OpenAI’s platform ~\cite{openai2023}, have further accelerated the creation of personalised chatbots and assistants. These models enable dynamic, context-aware conversations, enabling the design of chatbots that feel personal and adaptive to individual users.

By using prompts including customizable personalities, chatbots can assume different roles or characters that align with the user’s preferences. 
For example, Pataranutaporn et al.~\cite{pataranutaporn2021ai} explored how LLMs can be directed through prompt engineering to create supportive agents that serve as companions or mentors for education, self-reflection, or emotional support. This approach emphasises narrative identity and personalisation to enhance engagement and long-term impact. Their findings showed that AI characters can adapt to individual learning needs, increasing motivation, while also providing guidance for meditation and emotional support that contributes to mental well-being. 

While the aforementioned work indicates positive effects of plant care on people's well-being, we know little about how conversational AI can mediate and represent plants to positively influence interactions.

\section{Pilot Study and Application Evaluation}

The idea behind \textit{PlantWhisperer} was to explore how to foster personal, interactive connections between users and their plants. 
The application seeks to facilitate gardening and promote overall well-being through assistance from an AI chatbot. 
Next, we describe the pilot studies that informed the final application.

\begin{figure*}[t]
    \centering
    \includegraphics[width=1\linewidth]{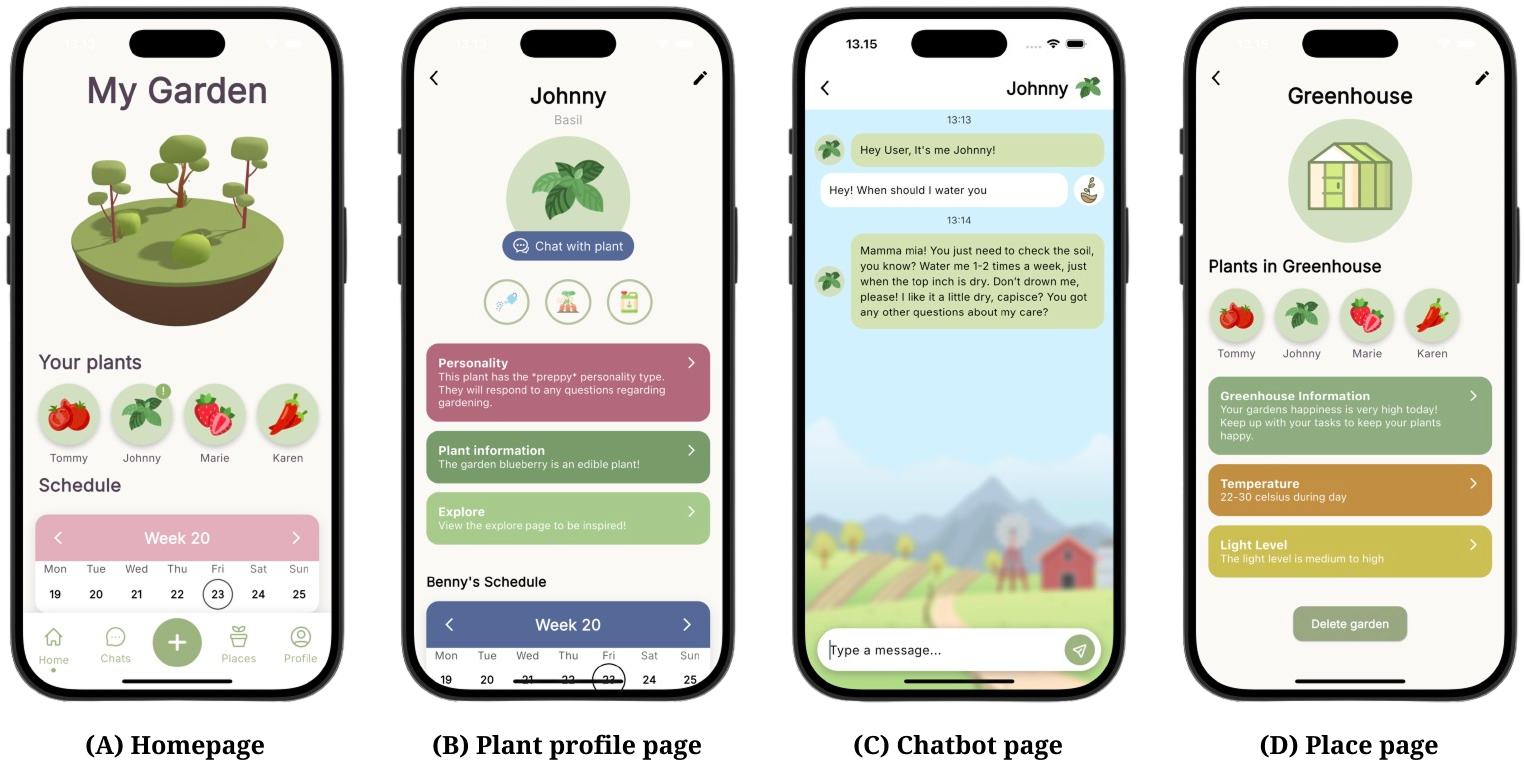}
    \caption{Screenshots of the PlantWhisperer application.}
    \Description{Four screenshots of the PlantWhisperer application. From left to right: homepage; plant profile page; chatbot page; and place page.}
    \label{fig:app_images}
\end{figure*}

\subsection{Piloting Application Design}
The application incorporates chatbots representing different plants, allowing users to communicate with their plants through natural language. 
The application consists of five main pages, all located in the navigation bar, as shown in \autoref{fig:app_images}. The different pages are designed to contain various contextual elements provided by the user. These contextual elements help the LLM provide appropriate support for plant care.

To ensure the quality and relevance of the responses generated by the plant chatbots, we conducted a round of prompt tests. The test involved four participants and four plant chatbots: Basil, Chilli, Tomato and Strawberry. The plants' behaviour was determined by three criteria: their visual appearance, their level of maintenance, and their culinary experience. 
The test consisted of two structured scenarios. In the scenarios, participants were asked to think aloud and imagine they had just downloaded the `PlantWhisperer' application and added a plant. They then interacted with the chatbot using predefined questions and subsequently searched for answers to those same questions on Google.
Both scenarios were repeated for the two different plants assigned to each participant.

Results from the test suggested that participants preferred the chatbot over Google for plant care, finding it more specific, easier to understand, and more engaging. One participant noted, \textit{`I don't have to search as much because it already understands it'}. Participants generally liked the plants having personalities, saying it made the interaction more enjoyable and felt like a real conversation. However, some found the chatbot’s responses too long, which negatively impacted their experience. Follow-up questions were appreciated for keeping the conversation natural and engaging, with one participant saying they \textit{`piqued my curiosity'}. 
Participants' feedback prompted refinements to balance the tone intensity and message length, while ensuring key contextual elements and follow-up questions were preserved.

\begin{table*}[h]
  \caption{Examples of participant quotes illustrating the main takeaways.}
  \label{tbl:examples}
  \centering
  \small
  \resizebox{\linewidth}{!} {
  \begin{tabular}{lp{0.7\linewidth}}
  \toprule
  \textbf{Takeaway} & \textbf{Example Participant Quote} \\
  \midrule
  (1) & ``\textit{The conversation kind of draws you in. I can
imagine that you’d want to find out more interesting
things.}'' (P7) \\
  
  (1) & ``\textit{I think it’s just that feeling of satisfaction, and now I have some herbs I can use.}'' (P1) \\
  
  (1) & ``\textit{A feeling of pride in yourself, that you’re actually capable of taking care of it.}'' (P3) \\

  (2) & ``\textit{That uncertainty can be really discouraging.
You try watering too little, then too much. You just
don’t know what you’re doing.}'' (P10)\\

  (2) & ``\textit{I learned that yellow leaves can indicate over-watering, and that if there is mold on your plant, you should increase air circulation.}'' (P3) \\
      
  (2) & ``\textit{I did not know anything about basil plants before because I have never grown one. So yes, I did, because it answered my questions.}'' (P4) \\
  \bottomrule 
  \end{tabular}}
\end{table*}

\subsection{Evaluation}\label{Evaluation}
We next recruited ten participants to evaluate a fully functional version of `PlantWhisperer'. The purpose of the evaluation was to examine how the participants interacted with the application and plant chatbots, and to build an understanding of PlantWhisperer's potential to support overall well-being. 
To ensure consistent data collection and support participants' immersion, we limited the study to a single plant chatbot.

Participants were provided with a phone running the application.
We placed a basil plant in front of them to create a realistic setting for growing basil. 
The interviews consisted of an introduction and two scenarios with five tasks and interview questions. One of the researchers briefly presented the application and the test procedure. Participants then created a user account, logged in, and browsed for two minutes before engaging with the first scenario.

We based the interview questions on the PERMA Model~\cite{195905820110101}, centred on well-being `pillars' of which we included three: 
`Positive Emotions', `Engagement', and `Accomplishment' (PEA). The model was used as a conceptual framework for structuring the interview guide to investigate participants’ subjective experiences related to PEA.
`Positive Emotions' refers to what people feel e.g. comfort and joy. `Engagement' refers to the state of flow where a person becomes so absorbed in an activity that they lose track of time and feelings. 
`Accomplishment' refers to the pursuit of mastery, achievement and success for its own sake, contributing to personal fulfilment and well-being.
For the full interview guide, see \autoref{Eval. Interview guide}.
We include the two scenarios below:

\pagebreak

\begin{itemize}
    \item \textbf{Scenario 1}:
We asked participants to ``\textit{imagine that you have planted a seed, and your plant is beginning to sprout. You want to use PlantWhisperer to learn more about how to take care of your plant}''.
The participants were subsequently guided through five tasks connected to the interaction between the plant chatbot and the participant. 
After completing the tasks, we interviewed participants about their experience with the application and their chats with the plant regarding \textbf{Positive Emotions} and \textbf{Engagement}.
All tasks and questions are listed in \autoref{Eval. Interview guide}.

\item \textbf{Scenario 2} was centred around \textbf{Accomplishment}, while also immersing the participants further into the mentality of growing their own plant. This part of the evaluation focused on interacting with a real basil plant while relating the experience to the app. At the end of Scenario 2, participants were asked questions about the general use of the application. The questions capture the overall experience with the application and with the plant chatbot. An example question: ``\textit{Do you feel that the app gives you sufficient amount of information about your plants and tasks?}''.
\end{itemize}

\section{Analysis and Takeaways}

To analyse the qualitative interview data, thematic analysis was employed, following the six-step approach outlined by Braun and Clarke ~\cite{braunandclarke2006}. This method was selected for its flexibility and its ability to identify and interpret patterns of meaning across the qualitative data. 
All interviews were audio recorded and subsequently transcribed using Whisper~\cite{whisper2023}. 
The analysis was carried out collaboratively by four authors using a Miro board.
In total, 168 meaningful quotes were selected from the ten interview transcriptions. During initial theme generation, related codes were grouped into overarching categories that captured recurring patterns in the data. Following several iterations, categories were refined into ten categories (labelled C1 to C10). The ten categories (\textit{1: Conversation flow and messages; 
2: Emotions, connection, and engagement;
3: Plant personality;
4: Relationship with plants;
5: App functions and UI;
6: App context;
7: Contrasting tools;
8: Learning;
9: Food and recipes;
10: Further development})
were subsequently grouped under three overarching themes (\textit{1: Chatting with plant chatbots;
2: Support and learning in plant care;
3: App context and interaction})
that broadly encompasses their content. 

The evaluation of \textit{PlantWhisperer} resulted in two main takeaways (see Table~\ref{tbl:examples} for illustrative quotes).

\textbf{(Takeaway 1) Plant chatbots can foster a connection with plants.}
Several participants appreciated the plant chatbot's playful and emotionally expressive personality, which made the interaction more enjoyable and meaningful. For these users, the conversational tone contributed to a stronger bond with their plants and even a sense of responsibility for their care. 
This suggests that emotionally rich design elements in conversational agents, such as plant chatbots, can influence user engagement and potentially support positive behavioural outcomes, such as sustained plant care.

\textbf{(Takeaway 2) Plant chatbots show potential for supporting learning}.
Participants highlighted and appreciated the plant chatbot's context-aware responses compared to web searches. However, preferences regarding chatbot output varied. Some participants preferred an emotionally engaging personality, while others favoured more direct and concise guidance. These findings suggest that flexibility in interaction style was appreciated, combined with the use of contextual and historical plant information.

\section{Discussion and Conclusion}
The evaluation of PlantWhisperer suggests that plant chatbots can positively influence plant care. 
Several participants stated that the conversation with the plant chatbot was engaging and easy to understand.
They expressed appreciation for the plant chatbot’s characteristics and playful tone, suggesting that the social and emotional aspects of the interaction enhanced their overall experience. In some cases, this contributed to a stronger sense of responsibility toward their plants. 
Conversely, other participants found the plant chatbot's expressive tone and verbose replies to be unnecessary or even distracting. These participants expressed that they would have preferred more information-focused responses, emphasising efficiency over engagement and connection. This occurred despite the fact that it had already been considered following the prompt testing, suggesting that the adjustments made at that stage were either insufficient or not fully aligned with user expectations during real interactions. It highlights the importance of ongoing evaluation and iteration to ensure that design choices effectively address user needs.
Taken together, these results show that users have different needs when chatting with the plant chatbot. 
Some users appreciate connecting with their plants, while others are more likely to use plant chatbots exclusively as gardening tools.

The results surrounding the plant chatbots' personalities and emotional connection, reveal the potential of affecting well-being through interacting and bonding when using conversational interfaces. Several participants linked plant care to feelings of achievement, connection, and even emotional regulation. These insights align with the growing research suggesting that digital tools, when thoughtfully designed, can foster emotional enrichment ~\cite{pataranutaporn2025oceanchateffectvirtualconversational}. 
The 'Caring for Vincent' chatbot~\cite{10.1145/3290605.3300932} and `PlantWhisperer' are different in regard to the distribution of care and guidance roles. For example, `PlantWhisperer' is simultaneously care-receiving and guidance-giving, whereas the user is care-giving and guidance-receiving. Additionally, the two concepts differ in their core purposes: Vincent is designed primarily to support users' mental health directly, whereas `PlantWhisperer' serves as a gardening assistant, with its potential to benefit well-being.
Additionally, PlantWhisperer's physical aspect differs from that of existing works.
Adding a plant as a physical aspect creates possibilities for further supporting users' well-being and engagement. 

Despite the contributions of this work, several limitations should be acknowledged. While a longitudinal evaluation of PlantWhisperer would have been preferable, practical constraints led us to conduct a controlled study. Ideally, this would have taken the form of a field study in which participants used the application in their own environments over an extended period. In contrast, controlled settings limit the emergence of insights that arise through everyday use in real-world contexts~\cite{fieldStudy}. A longitudinal field study would likely surface additional findings and support the development of deeper, more sustained relationships between users and plant chatbots.

For future work, an interesting direction surfaced by participants is to explore group chats. 
This could involve multiple plant chatbots and potentially other users, fostering connection with plants, learning about their care, and, in turn, supporting care for oneself.

\bibliographystyle{ACM-Reference-Format}
\bibliography{references}

\clearpage

\appendix

\section{Task and Interview Guide}\label{Eval. Interview guide}

\textbf{Scenario 1: }
Imagine that you’ve just planted a seed and made it sprout. You now want to use the PlantWhisperer app to learn more about how to take care of your plant. 
Please say out loud what you are typing, and what it replies.
\\
\textbf{A test with tasks that explore the chatbot:} The user makes their own account and gets to explore the app and click on the pages. The user should not begin a chat yet.

\begin{itemize}
    \item \textit{Task 1:}
Your plant has started to get yellow leaves, talk to the plant and see how to best fix this.
\item \textit{Task 2:}
You have a greenhouse space, which the chatbot knows about. Ask the chatbot where it is located. Try to find out what conditions are best for your plant given the location.
\item \textit{Task 3:}
Imagine you have just watered your plant. Log your watering in the calendar. 
Tell the plant that you have just watered it. 
\item \textit{Task 4: }
Imagine that your plant has started showing signs of mold. Ask your plant about your problem.
\item \textit{Task 5:}
You now have the opportunity to talk with the plant freely. Ask it anything you are curious about. 
\end{itemize}

\noindent\textbf{Semi-Structured interview on PEA: }
We’ll now ask some questions where we’d like to hear a bit about your experience with the app and chatting with the plant.
\\
\textit{Engagement:}
\begin{itemize}
    \item Do you feel engaged in the conversation? 
When chatting with the plant do you feel a connection to the plant? If yes, what sort of connection?
\item Could you imagine yourself talking with the plant about other things than just gardening? 
\item Was it hard/easy to focus and immerse yourself in the activity? 
\end{itemize}

\textit{Positive Emotions:}
\begin{itemize}
    \item How do you feel when chatting/texting with the plant?
\item Could you imagine yourself feeling a sense of mindfulness when taking care of your plant?
\item Do you think growing or caring for the plant could have an effect on your mood or emotional state in general? If so, how?
\end{itemize}

\noindent\textbf{Scenario 2: }
The participants will be given a plant they can touch, look at and so on. They will be asked to imagine that they have grown this plant from beginning to finish, meaning that they have watered, fertilized the plant, and dealt with eventual difficulties. They will be asked further questions:

\textit{Accomplishment: }
\begin{itemize}
    \item Do you feel as if you are developing skills when chatting with the plant? 
If yes: How does it make you feel? 
\item If you were to stand in front of your plant, observing it and the growth progress what would you feel? 
\item Can you describe whether activities such as water and fertilizing your plant gives you a feeling of accomplishment or progress?
\end{itemize}

\textit{General use of the application:}
\begin{itemize}
    \item Do you feel like the app provides sufficient information about your plant and tasks?
    \item Would you use this app again, to learn further about your own plant(s) at home? Why/Why not?
    \item How do you think PlantWhisperer differentiates from asking Google or other apps about your plant(s)? - If so, why?
    \item Was there any functionality you felt was missing in the app?
\end{itemize}

\end{document}